\documentclass[%
 reprint,
 amsmath,amssymb,
 aps,
pra,
]{revtex4-1}

\usepackage{graphicx}% Include figure files
\usepackage{dcolumn}% Align table columns on decimal point
\usepackage{bm, color}% bold math
\usepackage{algorithmic,color}
\usepackage{algorithm}

\def\>{\ensuremath{\rangle}}
\def\<{\ensuremath{\langle}}

\def\-{\ensuremath{\textrm{-}}}

\def\h{\ensuremath{\mathcal{H}}}

\def\l{\ensuremath{\mathcal{L}}}

\def\u{\ensuremath{\mathcal{U}}}
\def\k{\ensuremath{\mathcal{K}}}

\def\u{\ensuremath{\mathcal{U}}}

\def\x{\ensuremath{\mathcal{X}}}

\def\v{\ensuremath{\mathcal{V}}}

\def\a{\ensuremath{\mathcal{A}}}
\def\b{\ensuremath{\mathcal{B}}}

\def\e{\ensuremath{\mathcal{E}}}

\def\l{\ensuremath{\mathcal{L}}}

\def\d{\ensuremath{\mathcal{D}}}

\def\<{\langle}
\def\>{\rangle}
\def\l{\mathcal{L}}
\def\k{\mathcal{K}}

\newtheorem{theorem}{Theorem}

\newtheorem{corollary}{Corollary}
\newtheorem{lemma}{Lemma}

\newtheorem{definition}{Definition}

\begin{document}

\preprint{APS/123-QED}

\title{The Structure  of Decoherence-free Subsystems}% Force line breaks with \\

\author{Ji Guan}
\affiliation{Center for Quantum Software and Information,
University of Technology Sydney, NSW 2007, Australia }

%\altaffiliation[Also at ]{Center for Quantum Computation and Intelligent Systems,
%University of Technology Sydney, NSW 2007, Australia}
%\and
%State Key Laboratory of Intelligent Technology and Systems,\\ Department of Computer Science and Technology, Tsinghua University, Beijing 100084, China }%Lines break automatically or can be forced with \\
\author{Yuan Feng}%
%\author{Mingsheng Ying}
\affiliation{Center for Quantum Software and Information,
University of Technology Sydney, NSW 2007, Australia }
\author{Mingsheng Ying}
 \affiliation{Center for Quantum Software and Information,
University of Technology Sydney, NSW 2007, Australia}
\affiliation{Institute of Software, Chinese Academy of Sciences, Beijing 100190, China}
 \affiliation{Department of Computer Science and Technology, Tsinghua University, Beijing 100084, China}
 \date{\today}% It is always \today, today,

\begin{abstract}
Decoherence-free subsystems have been successfully developed as a tool to preserve fragile quantum information against noises.  In this letter, we develop a structure theory for decoherence-free subsystems. Based on it, we present an effective algorithm to construct a set of maximal decoherence-free subsystems in the sense that any other such subsystem is a subspace of one of them. As an application of these techniques in quantum many body systems, we propose a simple and numerically robust method to determine if two irreducible tensors are repeated, a key step in deciding if they are equivalent in generating matrix product states.
\end{abstract}

\pacs{Valid PACS appear here}% PACS, the Physics and Astronomy
\maketitle

%\tableofcontents

To build large scale quantum computers, the obstacles, such as decoherences and noises,  must be managed and overcome~\cite{nielsen2010quantum}. One of the effective methods for this purpose is through
decoherence-free subspaces proposed by Daniel A. Lidar in \cite{lidar1998decoherence}. A subspace of the system Hilbert space is said to be decoherence-free if the effect of the noise on it is simply unitary, and thus easily correctable. For this sake, decoherence-free subspaces are important tools in quantum computing, where coherent control of quantum systems is often the desired goal \cite{lidar2012review}. On the other hand, decoherence-free subspaces  can be characterized as a special case of quantum error correcting codes to preserve quantum information against noises \cite{lidar2012review}. Indeed, we do not even need to restrict the decoherence-free dynamics to a subspace. E. Knill, R. Laflamme, and L. Viola introduced  the concept of noiseless subsystems, by extending higher-dimensional irreducible representations of the algebra generating the dynamical symmetry in the system-environment interaction \cite{knill2000theory}. A subsystem is a factor in the tensor product decomposition of a subspace, and 
the noiseless subsystem requires the evolution on it to be identity. 

Noiseless subsystems have been fully characterized and intensely studied in \cite{choi2006method,blume2010information,beny2007generalization,kribs2006quantum,kribs2005unified}. Remarkably, a structure theory of noiseless subsystems  was established in \cite{choi2006method}, leading to an algorithm which finds all noiseless subsystems for a given quantum operation (i.e. the evolution of an open quantum system, mathematically modeled by a super-operator) \cite{knill2006protected,wang2013numerical}. For the more general case of decoherence-free subsystems, however, a structure theory is still lacking, although several conditions for their existence were found in \cite{shabani2005theory}, and subsystems with significantly reduced noises  were carefully examined in \cite{wang2016minimal}. Without such a structure theory, it is hard to compute all decoherence-free subsystems (subspaces) or the highest-dimensional ones for a given super-operator. 

The aim of this letter is to develop a structure theory that shows precisely how a super-operator determines its decoherence-free subsystems, with the structure theory of noiseless subsystems as a special case. As an application, we develop an algorithm  to generate a set of maximal decoherence-free subsystems for any given super-operator such that any other decoherence-free subsystem is a subspace of one of them. Furthermore, we use this structure theory in the quantum many-body system described by a family of matrix product states generated by a tensor and find a  feasible way to numerically derive a basis for the tensor. Such a basis plays an important role in establishing the fundamental theorems of matrix product states \cite{cirac2017matrix,cuevas2017irreducible}.

%This paper is organized as follows. We recall some basic notions of quantum information theory and, in particular, introduce a central concept, continuous coherence,  in Section \uppercase\expandafter{\romannumeral2}. In Section \uppercase\expandafter{\romannumeral3}, we review the structure theory of noiseless subsystems by studying the fixed points of super-operators. We then establish a corresponding structure theory for decoherence-free subsystems in Section \uppercase\expandafter{\romannumeral4}, which leads to an algorithm for constructing, with the assumption of perfect initialization, a set of maximal such subsystems  for any given super-operator. In Section \uppercase\expandafter{\romannumeral5}, we present a procedure for checking whether or not a subsystem with a co-subsystem is decoherence-free under imperfect initialization. Furthermore, in Section \uppercase\expandafter{\romannumeral6}, 
%we apply the results obtained in the previous sections to find a basis for any tensor, generating a set of matrix product states that represent a quantum many-body system. A brief conclusion is drawn in the last section.
 Recall that given a quantum system $S$ with the associated (finite-dimensional) state Hilbert space $\h$, the evolution of the system can be mathematically  modeled by a super-operator, i.e. a completely positive  and trace-preserving (CPTP) map $\e$ on $\h$.  We say that a quantum system $A$ is a subsystem of $S$ if $\h=(\h_A\otimes\h_B)\oplus (\h_A\otimes\h_B)^\perp$ for some co-subsystem $B$, where $\h_A$ and $\h_B$ are the state spaces  of $A$ and $B$, respectively. Generally, co-subsystem $B$ is not unique and one may construct other co-subsystems of $A$ as subspaces of $\h_B$, or by combining co-subsystems of $A$ with orthogonal supports. For any two Hilbert spaces $\h$ and $\h'$, let $\l(\h,\h')$ be the set of all linear operators from $\h$ to $\h'$. Simply, we let $\l(\h)=\l(\h,\h)$ and $\d(\h)$ be the set of all quantum states, i.e. density operators with unit trace, on $\h$.  The support of a quantum state $\rho$, denoted by supp$(\rho)$, is the linear span of the eigenvectors corresponding to non-zero eigenvalues of $\rho.$ 
 \begin{definition}\label{eq_dfs_def}
   Let $\e$ be a super-operator on $\h$. A subsystem 
   $\h_A$ of $\h$ is called decoherence-free if there is a co-subsystem $\h_B$ of $\h_A$ (that is, $\h_A\otimes\h_B$ is a subspace of $\h$) and a unitary matrix $U_A$ on $\h_A$ such that  $\forall \rho_A\in D(\h_A) ,\ \forall \rho_B\in D(\h_B)$,\begin{eqnarray}\label{Eq_def_U}
     \exists \sigma_B\in D(\h_B): \e(\rho_A\otimes \rho_B)=U_A\rho_AU_A^\dagger\otimes \sigma_B.
   \end{eqnarray}
  Furthermore, if $U_A=I_A$, the identity operator on $\h_A$, then we say that $\h_A$ is noiseless.  
\end{definition}

Obviously, co-subsystem $\h_B$ of decoherence-free subsystem $\h_A$ is inessential  as  $\h_B$ can be traced over.
  Noiseless subsystems  have been intensely studied in the areas of quantum error correction \cite{kribs2005operator,beny2007generalization,choi2006method,kribs2005unified} and quantum memory \cite{kuperberg2003capacity}, and can be characterized by the set of fixed points of $\e$, denoted by $fix(\e)=\{A\in \l(\h)|\e(A)=A\}$.  From \cite{blume2010information,wolf2012quantum,blume2008characterizing}, with an appropriately orthogonal decomposition of the Hilbert space $\h=\bigoplus_{k=1}^n(\h_{A_k}\otimes \h_{B_k})\oplus \mathcal{K}$,  $fix(\e)$ admits a useful structure: 
\begin{eqnarray}\label{eq_decomposition}
  fix(\e)=\bigoplus_{k=1}^{n}\left(\l(\h_{A_k}) \otimes \sigma_k\right) \oplus 0_\k
\end{eqnarray}
where $\sigma_k$ is  a full-rank quantum state on $\h_{B_k}$.  
This decomposition is unique and called the fixed-point decomposition of $\h$ and can be computed by applying the structure of  $C^*$-algebra generated by the Kraus operators of $\e$; see \cite{knill2006protected,guan2016decomposition,wang2013numerical} for details.  It is easy to see that for each $k$, $\h_{A_k}$ is a noiseless subsystem. Conversely, this decomposition captures all noiseless subsystems; that is, $\h_A$ is a noiseless subsystem if and only if $\h_A\subseteq \h_{A_k}$ for some $k$. 

\emph{Example.---}
  Given $\h=\h_A\otimes \h_B$, and $\{|k\rangle_A\}_{k=0}^3$ and $\{|k\rangle_B\}_{k=0}^2$ are orthonormal bases of $\h_A$ and $\h_B$, respectively, let $\e$ be a super-operator on $\h$ with the Kraus operators:
  \begin{eqnarray*}
    E_{1}&=&|00\rangle\langle01|+|10\rangle\langle11|-|20\rangle\langle21|-|30\rangle\langle31|\\
    E_{2}&=&|01\rangle\langle00|+|11\rangle\langle10|-|21\rangle\langle20|-|31\rangle\langle30|\\
    E_{3}&=&|00\rangle\langle02|+|10\rangle\langle12|-|20\rangle\langle22|-|30\rangle\langle32|
  \end{eqnarray*}
  where $|kl\rangle=|k\rangle_A\otimes |l\rangle_B$. It is easy to calculate the  fixed-point  decomposition of $\h$ as 
  $$\h=\bigoplus_{l=1}^2\left[\h_l\otimes \h'\right] \oplus\k$$
  where $\h_1=\textrm{lin.span}\{|0\rangle_A,|1\rangle_A\}$, $\h_2=\textrm{lin.span}\{|2\rangle_A,|3\rangle_A\}$, $\h'=\textrm{lin.span}\{|0\rangle_B,|1\rangle_B\}$, and $\k = \h_A\otimes \textrm{lin.span}\{|2\rangle_B\}$. Then we can store 1-qubit  quantum information in $\h_1$ or $\h_2$.

Now, let us see the structure of $fix(\e)$ in a different way.  For any $k$ and $|\psi_k\rangle\in \h_k$, $|\psi_k\rangle\langle\psi_k|\otimes \sigma_k$ is a minimal stationary state; a state  $\rho$ is  stationary if $\rho\in fix(\e)$, and  $\rho$  is further minimal if there is no other stationary state $\sigma$ such that $\textrm{supp}(\sigma)\subseteq \textrm{supp}(\rho).$ Moreover, the support of a minimal stationary state is called a minimal subspace. Then 
 we can decompose $\h$ into a set of mutually orthogonal minimal subspaces with the subspace $\k$: 
\begin{eqnarray}\label{eq_mini_dec}
  \h=\bigoplus_{p=1}^m\h_p\oplus\k.
\end{eqnarray}
It is worth noting that each $\h_p$ is invariant under $\e$, i.e. for any $A\in \l(\h_p)$, $\e(A)\in\l(\h_p)$. 
Thus, the Kraus operators $\{E_k\}$ of $\e$ have the  corresponding block form:
$$\renewcommand{\arraystretch}{1.2}
E_k=\left[\begin{array}{c|c}
  \begin{array}{cccc}
  E_{k,1} & & &\\
  & E_{k,2}  & &\\
  & & \ddots &\\
  & & & E_{k,m}  
  \end{array} & T_k\\
  \hline
0&K_k
\end{array}\right]$$
 for some operators $E_{k,p}\in \l(\h_p)$, $K_k\in \l(\k)$, and $T_k\in \l(\k,\k^{\perp}).$
We then define a set of associated maps $\{\e_{p,q} : p,q = 1, \dots, m\}$ of $\e$:
\begin{eqnarray}\label{Eq_ass_maps}
  \e_{p,q}(\cdot)=\sum_{k} E_{k,p}\cdot E_{k,q}^\dagger.
\end{eqnarray}
Obviously, for any $p$ and $q$, $\e_{p,q}$ is a linear  map from $\l(\h_q,\h_p)$ to itself. If $p\not =q$, $\l(\h_q, \h_p)$ can  be viewed as (outer) coherences from  $\h_q$ to $\h_p$, i.e.  upper off-diagonal blocks of all matrices restricted in the decomposition $\h_p\oplus\h_q.$ Thus the coherence between $\h_p$ and $\h_q$ is  $\l(\h_q, \h_p)\oplus\l(\h_p, \h_q)$ and $\l(\h_q)$ can be regarded as inner coherences.

For all $p$ and $q$, the following two properties are easy to observe:
\begin{enumerate}
\item $\l(\h_q,\h_p)$ is invariant under $\e$; that is, for all $A\in \l(\h_q,\h_p)$, $\e(A)\in \l(\h_q,\h_p)$.
\item  $\lambda(\e_{p,q})\subseteq \lambda(\e)$, where $\lambda(\cdot)$ is the set of eigenvalues of a linear map.
\end{enumerate}
Furthermore,  the coherence $\l(\h_q,\h_p)$ is said to be continuous if there exists $A\in \l(\h_q,\h_p)$ such that $\e(A)=\e_{p,q}(A)=e^{i\theta}A$ for some real number $\theta$; that is, $\lambda(\e_{p,q})$  has an element with magnitude one.  Specially, if $\theta=0$, then $\l(\h_q,\h_p)$ is stationary. Obviously, inner coherence $\l(\h_q)$ is always stationary because a super-operator has at least one stationary state. Stationary coherences have been intensely studied in~\cite{baumgartner2012structure}, and can be easily checked.
\begin{lemma}[\cite{baumgartner2012structure}]\label{Lem_SC}
  Let $\e$ be a super-operator on $\h$ with the orthogonal decomposition presented in Eq. (\ref{eq_mini_dec}). Then for any $1\leq p, q\leq m$, $\l(\h_p,\h_q)$ is stationary if and only if there is a unitary matrix $U$ such that 
  $E_{k,p}=UE_{k,q}U^\dagger$ for all $k$.
  %, where $\{E_{k,p}\}_k$ and $\{E_{k,q}\}_k$ are the restriction of  Kraus operators of $\e$ onto $\h_p$ and $\h_q$, respectively.  
  Furthermore, $\h_p\simeq\h_q.$
\end{lemma}
Meanwhile, through studying stationary coherences,  the authors also obtained  the same structure of $fix(\e)$ in Eq.(\ref{eq_decomposition}). 
In the following discussion, we will show that continuous coherences imply the structure of decoherence-free subsystems.

\emph{Structure theorem.---}
By the definition, a decoherence-free subsystem $\h_A$  is a small section of the whole state space $\h$ in which the effect of the quantum noise modeled by $\e$ is equivalent to a unitary transformation. From Eq. (\ref{Eq_def_U}), the restriction of $\e$ onto $\h_A\otimes\h_B $, where $\h_B$ is the co-subsystem of $\h_A$, can be written as \begin{eqnarray}\label{eq_dfs_form}
  \e_{AB}=\u_A\otimes \e_{B}
\end{eqnarray} where $\u_A$ is a unitary super-operator on $\h_A$ and $\e_B$ is a super-operator on $\h_{B}.$  By the decomposition Eq.(\ref{eq_mini_dec}), $\h_B$ can be chosen to be  irreducible, i.e. the whole space $\h_B$ is minimal under $\e_B$. It is easy to observe from Eq.(\ref{eq_decomposition}) that $\e$ is irreducible if and only if there is only one stationary state and it is full-rank. From now on, we assume without loss of generality that the co-subsystem of a decoherence-free subsystem is always irreducible. 

First, we observe that  the joint system of a decoherence-free subsystem and its irreducible co-subsystem can be decomposed into minimal subspaces with continuous coherences. 
\begin{theorem}\label{Theo_dfs_cc}
    Given a super-operator $\e$ on 
    $$\h=(\h_A\otimes\h_B)\oplus (\h_A\otimes\h_B)^\perp .$$ 
    Let $\h_A$ be a decoherence-free subsystem and $U_A$ the corresponding unitary matrix in Eq.(\ref{Eq_def_U}). If  $\{|x\rangle\}_{x=1}^{m}$ is a set of mutually orthogonal eigenvectors of $U_A$ and $\h_x=\textrm{lin.span}\{|x\rangle\}\otimes \h_B$, then for all $1\leq p,q\leq m$,  $\h_p$  is a minimal subspace and
    $\l(\h_p,\h_q)$ is continuous.
  \end{theorem}  
{\it Proof.} Note that we assume $\h_B$ is irreducible. Let $\rho$ be the unique stationary state of $\e_B$.  Then for any $x$, $|x\rangle\langle x|\otimes \rho$ is a minimal stationary state of $\e$, and hence $\h_x$ is minimal. Furthermore, note that $U_A|x\rangle=e^{i\theta_x}|x\rangle$ for some $\theta_x$. Thus $$\e(|p\rangle\langle q|\otimes \rho)=e^{i(\theta_p-\theta_q)}|p\rangle\langle q|\otimes \rho$$ for all $p$ and $q.$
\hfill $\Box$

Theorem~\ref{Theo_dfs_cc} indicates that minimal subspaces with continuous coherences play an important role in determining decoherence-free subsystems. To check if two orthogonal minimal subspaces have a continuous coherence, we present the following lemma.

\begin{lemma}\label{lem_block_eq}
  Let $\e$ be a super-operator on $\h$ with the orthogonal decomposition presented in Eq. (\ref{eq_mini_dec}). Then for any $1\leq p, q\leq m$, $\l(\h_p,\h_q)$ is continuous if and only if there is a unitary matrix $U$ and a real number $\theta$ such that 
  $E_{k,p}=e^{i\theta}UE_{k,q}U^\dagger$ for all $k$.
  %, where $\{E_{k,p}\}_k$ and $\{E_{k,q}\}_k$ are the restriction of  Kraus operators of $\e$ onto $\h_p$ and $\h_q$, respectively.  
  Furthermore, $\h_p\simeq\h_q.$ 
\end{lemma}
{\it Proof.} 
Assume that $\l(\h_p,\h_q)$ is continuous; that is,  there is a matrix $A\in \l(\h_p,\h_q)$ such that $\e(A)=e^{i\theta}A$ for some real number $\theta$. Let $V=e^{-i\theta}P_q+I-P_q$, where $P_x$ is the projector onto $\h_x$,  and  $\v\circ\e(A)=A$ with $\v(\cdot)=V\cdot V^\dagger$. Moreover, it is obvious that $\h_p$ and $\h_q$ are also orthogonal minimal subspaces under $\v\circ\e$  by the decomposition Eq.(\ref{eq_mini_dec}). Therefore, there is a stationary coherence from $\h_q$ to $\h_p$ under $\v\circ\e.$ From Lemma~\ref{Lem_SC}, we have $\h_p\simeq\h_q$, and there exists some unitary matrix $U$ such that for any $k$,
$$E_{k,p}=e^{i\theta}UE_{k,q}U^\dagger.$$ 

Conversely, for any $p$ and $q$, let $$\e_{q,p}(\cdot)=\sum_{k}E_{k,q}\cdot E_{k,p}^\dagger=\sum_{k}e^{-i\theta}E_{k,q}\cdot UE_{k,q}^\dagger U^\dagger.$$ 
Its matrix representation \cite{guan2016decomposition} reads
\begin{eqnarray*}
  M_{q,p}&=&\sum_k e^{-i\theta}E_{k,q}\otimes (UE_{k,q}U^\dagger)^{*}\\
  &=&e^{-i\theta}(I\otimes U^{*})\left(\sum_k E_{k,q}\otimes E_{k,q}^{*}\right)(I\otimes U^{*^\dagger}).
\end{eqnarray*}
As $\lambda(M_{q,p})=\lambda(\e_{q,p})$ and $\sum_k E_{k,q}\otimes E_{k,q}^{*}$ is the matrix representation of $\e_{q,q}$ which is a super-operator and has 1 as one of its eigenvalues, we have $e^{-i\theta}\in\lambda(\e_{q,p}).$  
\hfill $\Box$

%The above lemma is a generalization of Lemma~\ref{lem_block_eq} in which stationary coherences need the Kraus operators are one-by-one unitarily equivalent. 
\begin{corollary}\label{cor_cc_eq}
Let $\e$ be a super-operator on $\h$ with the orthogonal decomposition presented in Eq. (\ref{eq_mini_dec}). 
Then the relation 
$\{(p, q) : 1\leq p, q\leq m,  \l(\h_p,\h_q)  \mbox{ is continuous}\}$
 is an equivalence relation. That is,
  for any $p$, $q$, and $r$, \begin{itemize}
    \item [(1)] (reflexivity) $\l(\h_p,\h_p)$ is continuous;
    \item [(2)] (symmetry) if $\l(\h_p,\h_q)$ is continuous, then so is $\l(\h_q,\h_p)$;
    \item [(3)] (transitivity) if $\l(\h_p,\h_q)$ and $\l(\h_q,\h_r)$ are both continuous, then so is $\l(\h_p,\h_r)$.
      \end{itemize}
\end{corollary}

With Corollary~\ref{cor_cc_eq}, $\l(\h_p,\h_q)$ is continuous coherence if and only if so is $\l(\h_q,\h_p)$. Thus in the following, we simply say that there is a continuous  coherence between $\h_p$ and $\h_q$ without referring to the direction. Then we group together minimal subspaces by continuous coherences.

\begin{theorem}\label{Theo_CC_dec}
  Let $\e$ be a super-operator on $\h$. There is a unique  orthogonal decomposition of $\h$
  \begin{eqnarray}\label{Eq_unique}
    \h=\bigoplus_{l=1}^{L}\x_l\oplus \k.
  \end{eqnarray}
  where \begin{itemize}
    \item[(1)] each $\x_l$ is either a minimal subspace or can be further decomposed into mutually orthogonal minimal subspaces with continuous coherences between any two of them:
    \begin{eqnarray}\label{Eq_dec_CC}
      \x_l=\bigoplus_{p=1}^{m_l}\b_{l,p}\simeq\mathbb{C}^{m_l}\otimes \b_l,  \ \ \b_l\simeq \b_{l,p} \ \ \forall p
    \end{eqnarray}  such that  the Kraus operators $\{E_k\}$ of $\e$ have a corresponding block form:
      \begin{eqnarray}\label{Eq_CC_dec_Kraus}
    \renewcommand{\arraystretch}{1.2}
E_k\simeq\left[\begin{array}{c|c}
  \begin{array}{ccc}
  U_{1}\otimes E_{k,1} &  &\\
   & \ddots &\\
   & & U_{L}\otimes E_{k,L}  
  \end{array} & T_k\\
  \hline
0&K_k
\end{array}\right]
  \end{eqnarray}
for some operators $E_{k,l}\in \l(\b_l)$, $K_k\in \l(\k)$, $T_k\in\l(\k,\k^\perp)$, and    unitary matrix $U_l=diag(e^{i\theta_{l,1}},\cdots,e^{i\theta_{l,m_l}})$ for some real numbers $\{\theta_{l,p}\}_{p=1}^{m_l}$ on $\mathbb{C}^{m_l}$. Moreover, $\b_l$ is irreducible under $\e_{l}(\cdot)=\sum_{k}E_{k,l}\cdot E_{k,l}^\dagger$. 
 Furthermore, 
$$fix(\e)=\bigoplus_l[fix(\u_{l})\otimes \rho_l]\oplus 0_\k,$$ 
where $\u_l(\cdot)=U_l\cdot U_l^\dagger$. 

    \item[(2)] there is no continuous coherence between any minimal subspaces $\b_{l, p}$ and $\b_{l', p'}$ whenever $l\neq l'.$ 
      \end{itemize}
\end{theorem}
{\it Proof.} See the Appendix. \hfill $\Box$

A similar decomposition of Eq.(\ref{Eq_dec_CC}) for only considering stationary coherences has been presented in \cite{baumgartner2012structure} and coincides with the fixed-point decomposition in Eq.(\ref{eq_decomposition}). We also review this in Appendix.
\begin{corollary}\label{Cor_mini}
  Let $\e$ be a super-operator  on $\h$ with the unique decomposition
$$\h=\bigoplus_l(\mathbb{C}^{m_l}\otimes \b_l)\oplus \k$$
presented in Theorem~\ref{Theo_CC_dec}. For any minimal subspace $\h'$, there is a pure state $|\psi\rangle\in \mathbb{C}^{m_l}$ for some $l$ such that $\h'=\textrm{lin.span}\{|\psi\rangle\}\otimes \b_l.$
\end{corollary}

%Let $A\in\b(\h)$ and $\e(A)=e^{i\theta}A$ for some $\theta$. Then $A\in B(\k^\perp)$ by \cite[Lemma 2]{guan2017super}. From the definition of continuous coherences, we have that $A=\bigoplus_l \bar{A}_l$ with $\in B(\mathbb{C}^m_l\bigotimes \b_l)$,  and $\sum_{k}U_l\otimes E_{k,l} \bar{A}_l U_l^\dagger\otimes E_{k,l}^\dagger=e^{i\theta}\bar{A}_l$. As $U_l$ is diagonal, $\bar{A}_l=$ 

 The above theorem shows  that minimal subspaces with continuous coherences can  be used to construct decoherence-free subsystem $\mathbb{C}^{m_l}$. On the other hand, we can show that other decoherence-free subsystems actually are all subspaces of the ones constructed in Eq.$(\ref{Eq_dec_CC})$.
 \begin{theorem}\label{Theo_find_DFS}
   Let $\e$ be a super-operator  on $\h$ with the unique decomposition:
$$\h=\bigoplus_l(\mathbb{C}^{m_l}\otimes \b_l)\oplus \k$$
presented in Theorem~\ref{Theo_CC_dec}. Then a
    subsystem $\h_A$ is decoherence-free if and only if \begin{itemize}
      \item [(1)] $\h_A\subseteq\mathbb{C}^{m_l}$ for some $l$, and
      \item [(2)] $\h_A$ is the support of some stationary state of $\u_l(\cdot)=U_{l}\cdot U_l^\dagger$,  where $U_l$ is the corresponding unitary matrix on $\mathbb{C}^{m_l}$ in the decomposition Eq.(\ref{Eq_CC_dec_Kraus}).  
    \end{itemize}
 
 \end{theorem}
 {\it Proof.} Assume that $\h_A$ is decoherence-free. By Theorem~\ref{Theo_dfs_cc}, Theorem~\ref{Theo_CC_dec} and Corollary~\ref{Cor_mini}, $\h_A\subseteq \mathbb{C}^{m_l}$ for some $l.$ From the definition of decoherence-free subsystems and the fact that the restriction of $\e$ onto $\mathbb{C}^{m_l}$ is  $\u_l$, $\h_A$ is a decoherence-free subspace under $\u_l$ and $\u_l(P_A)=P_A$, where $P_A$ be the projector onto $\h_A$. 

To prove the opposite direction,  we observe that if  $\h_A$ is the support of some stationary state of $\u_l$, then $P_AU_{l}P_A=P_AU_l=U_lP_A$. Thus $\h_A$ is a decoherence-free subspace under $\u_l$. The rest of the proof is direct from Theorem~\ref{Theo_CC_dec}. $\hfill$ $\Box$

This  theorem confirms that the set of decoherence-free subsystems $\{\mathbb{C}^{m_l}\}_l$ identified in Theorem~\ref{Theo_CC_dec} is  optimal; that is, any other decoherence-free subsystem is a subspace of one of them. So we only need to implement the decomposition in Theorem~\ref{Theo_CC_dec} and all decoherence-free subsystems can  be easily found by Theorem~\ref{Theo_find_DFS}.

One easy way of achieving this is to first transform all continuous coherences to stationary ones without changing any minimal subspace, and then use the  algorithm of the decomposition Eq.(\ref{eq_decomposition}) for stationary coherences, already proposed in the literature~\cite{guan2016decomposition,knill2006protected,wang2013numerical}. To be specific, for any two operators $E_{k,p}$ and $E_{k,q}$ in Eq. (\ref{Eq_CC_dec_Kraus}) of Theorem~\ref{Theo_CC_dec}, if they are unitarily equivalent with a phase $\theta$, i.e. $E_{k,p}\simeq e^{i\theta}E_{k,q}$, then 
 let $E'_{k,q}=e^{i\theta}E_{k,q}$ and $E'_{k,p}=E_{k,p}$. Note that $E'_{k,p}$ is unitarily equivalent to $E'_{k,q}$. The continuous coherence between $\h_p$ and $\h_q$ is transformed to be stationary, and $\h_p$ and $\h_q$ are still minimal subspaces under the new super-operator. Using this technique, we can  develop an algorithm to implement the decomposition in Theorem~\ref{Theo_CC_dec}.

Now we return back to see the example. By Theorem~\ref{Theo_CC_dec},   we can confirm that the first subsystem $\h_A\simeq\mathbb{C}^4$ is decoherence-free and further show that the evolution on it is a unitary operator $|0\rangle\langle0|+|1\rangle\langle1|-|2\rangle\langle2|-|3\rangle\langle3|$. Thus we can store  2-qubit information in this subsystem, which doubles the capacity of  noiseless subsystems.

\emph{Application: Matrix Product States.---}
The traditional techniques for describing quantum many body systems are usually not scalable due to the exponential growth of the dimension of the state space with the number of subsystems.  Matrix Product States (MPS), a special case of tensor networks (a theoretical and numerical tool describing quantum many-body systems), have been proved to be a useful family of quantum states for the description of  ground states of  one-dimensional quantum many body systems \cite{cirac2017matrix}.  

Given a  tensor $\a=\{A_{k}\in \mathcal{M}_D\}_{k=1}^{d}$ with a Hilbert space $\h_d=\textrm{lin.span}\{|k\rangle\}_{k=1}^{d}$, where $\mathcal{M}_{D}$ denotes the set of all $D\times D$ complex matrices, it generates a family of translationally invariant MPS, namely
$$V(\a)=\{|V_{n}(\a)\rangle\}_{n\in \mathbb{N^+}},$$
where $$|V_{n}(\a)\rangle=\sum_{k_1,\cdots,k_n=1}^{d}\textrm{tr}(A_{k_1}\cdots A_{k_n})|k_1\cdots k_{n}\rangle\in\h_d^{\otimes n}.$$
Here, each $|V_{n}(\a)\rangle$ corresponds to a state of $n$ spins of physical dimension $d.$
Let the associated completely positive map  be $\e_\a(\cdot)=\sum_{k=1}^{d}A_{k}\cdot A_{k}^\dagger$.

By \cite{cuevas2017irreducible}, we can always find a set of irreducible tensors $\{\a_{j}\}_{j=1}^{m}$ with the same Hilbert space  $\h_d$, and a set of complex number $\{\mu_{j}\}_{j=1}^{m}$ such that for any $n\in\mathbb{N^{+}}$,
\begin{eqnarray}\label{Eq_MPS}
  |V_n(\a)\rangle=\sum_{j=1}^{m}\mu_j^{n}|V_n(\a_{j})\rangle
\end{eqnarray}
where a tenor is called irreducible if the associated map is CPTP and irreducible. That is, for any tensor $\a$,  the generated MPS can be linearly represented by MPS of a set of irreducible tensors. Furthermore, we can identify irreducible tensors that are essentially the same in the following sense.
\begin{definition}[\cite{cuevas2017irreducible}]
  We say that two irreducible tensors $\a=\{A_{k}\}_{k=1}^{d}$ and $\b=\{B_{k}\}_{k=1}^{d}$ are repeated if there exist a phase $\theta$ and a unitary matrix $U$ such that 
  $$A_{k}=e^{i\theta}UB_kU^\dagger, \ \forall k.$$
\end{definition}

By the definition, if $\a$ and $\b$ are repeated, then $|V_n(\a)\rangle=e^{in\theta}|V_n(\b)\rangle$ for all $n\in \mathbb{N^+}$. Therefore, for any tensor $\a$, we can assume without loss of generality that the set of irreducible tensors $\{\a_{j}\}_{j=1}^{m}$ in Eq. (\ref{Eq_MPS}) are non-repeated. Such a set is called a basis of $\a$. 

To determine if two irreducible tensors are repeated, an obvious way is to work out the Jordan decomposition of all the matrices involved. However, as Jordan decomposition is sensitive to computational errors, it is not suitable for numerical analysis. Here we propose a more robust method to achieve this by using the results of continuous coherences. 
\begin{theorem}\label{thm:repeated}
  Let $\a=\{A_{k}\}_{k=1}^{d}$ and $\b=\{B_{k}\}_{k=1}^{d}$ be two irreducible tensors. Then they are repeated if and only if $\e_{\a,\b}$ has an eigenvalue with magnitude one, where $\e_{\a,\b}=\sum_{k=1}^d A_{k}\cdot B_{k}^\dagger.$
\end{theorem}
{\it Proof.} Let $A_k\in \l(\h_\a)$ and $B_k\in \l(\h_\b)$ for all $k$, and $\e$ be a super-operator on $\h_\a\oplus \h_\b$ with Kraus operators $\{diag(A_k,B_k)\}_{k=1}^d$. Obviously, $\h_\a$ and $\h_\b$ are both minimal subspaces under $\e$. Then the result follows directly from Lemma~\ref{lem_block_eq}. 
\hfill $\Box$

Note that many interesting results obtained for MPS rely on the basis. For example, one of the most fundamental problems of quantum many body systems is to identify different tensors that give rise to the same MPS.
%: for any two different tensors $\a$ and $\b$, they can generate the same MPS, i.e. $V(\a)=V(\b)$, which introduces an ambiguity for analyzing many-body states by MPS generated by tensors. 
This problem can be reduced to deciding if the bases of two given tensors are related by a unitary transformation~\cite{cuevas2017irreducible}. 
Thus Theorem~\ref{thm:repeated}, which employs simple linear algebra calculation to check if two irreducible tensors are repeated, a key step in constructing the bases of 
tensors, can help solve these problems.

\emph{Conclusion.---}
In this letter, we established a structure theory for decoherence-free subsystems. Consequently, a method for finding a set of maximal decoherence-free subsystems has been found.  As an application in many body quantum systems, these results give us a numerically robust way to find a basis for any tensor by computing the eigenvalues of some linear maps. 

For future studies, an immediate topic is to generalize our results to continuous-time quantum systems. In \cite{ticozzi2008quantum}, it was studied in the quantum control setting and the authors expected to obtain a linear-algebraic approach for finding all decoherence-free subsystems  for any given continuous-time quantum system. 
\section*{appendix}
\subsection*{The structure of  Noiseless Subsystems}
Noiseless subsystems are a special case of decoherence-free subsystems and have been intensely studied in the areas of quantum error correction \cite{kribs2005operator,beny2007generalization} and quantum memory \cite{kuperberg2003capacity}.  We first review the structure of  noiseless subsystems which can be characterized by minimal subspaces with stationary coherences.  

  Let $\e$ be a super-operator on $\h$. Then there is a unique  orthogonal decomposition of $\h$
  \begin{eqnarray}\label{Eq_NS_dec_unique}
    \h=\bigoplus_{l=1}^{L}\x_l\oplus \k
  \end{eqnarray}
  where: \begin{itemize}
    \item[(1)] each $\x_l$ is either a minimal subspace or can be further decomposed into mutually orthogonal minimal subspaces with stationary coherences between any two of them:
    \begin{eqnarray}\label{Eq_NS_dec}
      \x_l=\bigoplus_{p=1}^{m_l}\b_{l,p}\simeq\mathbb{C}^{m_l}\otimes \b_l, \ \ \b_l\simeq \b_{l,p} \ \forall p
    \end{eqnarray} 
    so that the Kraus operators $\{E_k\}$ of $\e$ have a block form
  \begin{eqnarray}\label{Eq_SC_dec_Kraus}
    \renewcommand{\arraystretch}{1.2}
E_k\simeq \left[\begin{array}{c|c}
  \begin{array}{ccc}
  I_{1}\otimes E_{k,1} &  &\\
   & \ddots &\\
   & & I_{L}\otimes E_{k,L}  
  \end{array} & T_k\\
  \hline
0&K_k
\end{array}\right]
  \end{eqnarray}
  in the corresponding basis, 
for some operators $E_{k,l}\in \l(\b_l)$, $K_k\in \l(\k)$, and $T_k\in\l(\k,\k^\perp)$. Here  $I_{l}$ is the identity operator on $\mathbb{C}^{m_l}$ and $\b_l$ is irreducible under $\e_{l}(\cdot)=\sum_{k}E_{k,l}\cdot E_{k,l}^\dagger$. Furthermore, 
$$fix(\e)\simeq\bigoplus_{l}[\l(\mathbb{C}^{m_l})\otimes \rho_{l}]\oplus 0_\k $$
    where $\rho_l$ is the unique stationary state of $\e_{l}$, and $0_\k$ is the zero operator on $\k$.
    \item[(2)] there is no stationary coherence between any minimal subspaces $\b_{l, p}$ and $\b_{l', p'}$ whenever $l\neq l'.$  
      \end{itemize}

\subsection*{The Proof of Theorem 2}

{\it Proof.} By the structure of noiseless subsystems, there is a unique  orthogonal decomposition of $\h$ as
  $$\h=\bigoplus_{l=1}^{L'}\x'_l\oplus \k$$
such that for any orthogonal minimal subspaces $\h_1$ and $\h_2$, they have stationary coherences if and only if $\h_{1}\oplus \h_2\in \x'_l$ for some $l$.
      Then we divide each $\{\x'_l\}$ into a finite number of disjoint subsets by continuous coherences; that  is for any $l_1\not =l_2$, if there is a continuous coherence between any minimal subspaces in $\x'_{l_1}$ and $\x'_{l_2}$, then they are in the same subset. This can be done as the existence of continuous coherences is an equivalence relation by Corollary~1. Then we define $\{\x_l\}_{l=1}^{L}$ to be the set of the direct sum  of all elements in each subset.  Therefore, $\h$ can be uniquely decomposed as $\h=\bigoplus_l\x_l\oplus \k.$ Obviously, for any two orthogonal  minimal subspaces $\b_{l_1}\in\x_{l_1}$ and $\b_{l_2}\in\x_{l_2}$, $\l(\b_{l_1},\b_{l_2})$ is continuous if and only if $l_1=l_2$.

Furthermore, for each $l$, $\x_l$ can be further decomposed to mutually orthogonal minimal subspaces:
$$\x_l=\bigoplus_{p=1}^{m_l}\b_{l,p}.$$
By Lemma~2, %we have that $\{\e_{l,p}\}_p$ is a set of quantum operation which are unitarily equivalent to each other. So 
in  an appropriate decomposition  of $\x_l=\bigoplus_{p=1}^{m_l}\b_{l,p}\simeq\mathbb{C}^{m_l}\otimes \b_l$ and $\b_l\simeq \b_{l,p}$ for all $p$:
\begin{eqnarray}
    \renewcommand{\arraystretch}{1.2}
E_k=\left[\begin{array}{c|c}
  \begin{array}{ccc}
  U_{1}\otimes E_{k,1} &  &\\
   & \ddots &\\
   & & U_{L}\otimes E_{k,L}  
  \end{array} & T_k\\
  \hline
0&K_k
\end{array}\right]
  \end{eqnarray}
and $\b_l$ is irreducible under $\e_{l}(\cdot):=\sum_{k}E_{k,l}\cdot E_{k,l}^\dagger$  for all $l$, 
where $U_l=diag(e^{i\theta_{l,1}},\cdots,e^{i\theta_{l,m_l}})$ for a set of real numbers $\{\theta_{l,p}\}_{p=1}^{m_l}.$  From the structure of noiseless subsystems and noting that the stationary coherence is continuous, we have
$$fix(\e)=\bigoplus_l[fix(\u_{l})\otimes \rho_l]\oplus 0_\k,$$ 
where $\rho_l$ is the unique stationary state of $\e_l.$ \hfill $\Box$
\\
\\
\bibliography{bib}

\begin{thebibliography}{22}%
\makeatletter
\providecommand \@ifxundefined [1]{%
 \@ifx{#1\undefined}
}%
\providecommand \@ifnum [1]{%
 \ifnum #1\expandafter \@firstoftwo
 \else \expandafter \@secondoftwo
 \fi
}%
\providecommand \@ifx [1]{%
 \ifx #1\expandafter \@firstoftwo
 \else \expandafter \@secondoftwo
 \fi
}%
\providecommand \natexlab [1]{#1}%
\providecommand \enquote  [1]{``#1''}%
\providecommand \bibnamefont  [1]{#1}%
\providecommand \bibfnamefont [1]{#1}%
\providecommand \citenamefont [1]{#1}%
\providecommand \href@noop [0]{\@secondoftwo}%
\providecommand \href [0]{\begingroup \@sanitize@url \@href}%
\providecommand \@href[1]{\@@startlink{#1}\@@href}%
\providecommand \@@href[1]{\endgroup#1\@@endlink}%
\providecommand \@sanitize@url [0]{\catcode `\\12\catcode `\$12\catcode
  `\&12\catcode `\#12\catcode `\^12\catcode `\_12\catcode `\%12\relax}%
\providecommand \@@startlink[1]{}%
\providecommand \@@endlink[0]{}%
\providecommand \url  [0]{\begingroup\@sanitize@url \@url }%
\providecommand \@url [1]{\endgroup\@href {#1}{\urlprefix }}%
\providecommand \urlprefix  [0]{URL }%
\providecommand \Eprint [0]{\href }%
\providecommand \doibase [0]{http://dx.doi.org/}%
\providecommand \selectlanguage [0]{\@gobble}%
\providecommand \bibinfo  [0]{\@secondoftwo}%
\providecommand \bibfield  [0]{\@secondoftwo}%
\providecommand \translation [1]{[#1]}%
\providecommand \BibitemOpen [0]{}%
\providecommand \bibitemStop [0]{}%
\providecommand \bibitemNoStop [0]{.\EOS\space}%
\providecommand \EOS [0]{\spacefactor3000\relax}%
\providecommand \BibitemShut  [1]{\csname bibitem#1\endcsname}%
\let\auto@bib@innerbib\@empty
%</preamble>
\bibitem [{\citenamefont {Nielsen}\ and\ \citenamefont
  {Chuang}(2010)}]{nielsen2010quantum}%
  \BibitemOpen
  \bibfield  {author} {\bibinfo {author} {\bibfnamefont {M.~A.}\ \bibnamefont
  {Nielsen}}\ and\ \bibinfo {author} {\bibfnamefont {I.~L.}\ \bibnamefont
  {Chuang}},\ }\href@noop {} {\emph {\bibinfo {title} {Quantum computation and
  quantum information}}}\ (\bibinfo  {publisher} {Cambridge university press},\
  \bibinfo {year} {2010})\BibitemShut {NoStop}%
\bibitem [{\citenamefont {Lidar}\ \emph {et~al.}(1998)\citenamefont {Lidar},
  \citenamefont {Chuang},\ and\ \citenamefont {Whaley}}]{lidar1998decoherence}%
  \BibitemOpen
  \bibfield  {author} {\bibinfo {author} {\bibfnamefont {D.~A.}\ \bibnamefont
  {Lidar}}, \bibinfo {author} {\bibfnamefont {I.~L.}\ \bibnamefont {Chuang}}, \
  and\ \bibinfo {author} {\bibfnamefont {K.~B.}\ \bibnamefont {Whaley}},\
  }\href@noop {} {\bibfield  {journal} {\bibinfo  {journal} {Physical Review
  Letters}\ }\textbf {\bibinfo {volume} {81}},\ \bibinfo {pages} {2594}
  (\bibinfo {year} {1998})}\BibitemShut {NoStop}%
\bibitem [{\citenamefont {Lidar}(2012)}]{lidar2012review}%
  \BibitemOpen
  \bibfield  {author} {\bibinfo {author} {\bibfnamefont {D.~A.}\ \bibnamefont
  {Lidar}},\ }\href@noop {} {\bibfield  {journal} {\bibinfo  {journal} {arXiv
  preprint arXiv:1208.5791}\ } (\bibinfo {year} {2012})}\BibitemShut {NoStop}%
\bibitem [{\citenamefont {Knill}\ \emph {et~al.}(2000)\citenamefont {Knill},
  \citenamefont {Laflamme},\ and\ \citenamefont {Viola}}]{knill2000theory}%
  \BibitemOpen
  \bibfield  {author} {\bibinfo {author} {\bibfnamefont {E.}~\bibnamefont
  {Knill}}, \bibinfo {author} {\bibfnamefont {R.}~\bibnamefont {Laflamme}}, \
  and\ \bibinfo {author} {\bibfnamefont {L.}~\bibnamefont {Viola}},\
  }\href@noop {} {\bibfield  {journal} {\bibinfo  {journal} {Physical Review
  Letters}\ }\textbf {\bibinfo {volume} {84}},\ \bibinfo {pages} {2525}
  (\bibinfo {year} {2000})}\BibitemShut {NoStop}%
\bibitem [{\citenamefont {Choi}\ and\ \citenamefont
  {Kribs}(2006)}]{choi2006method}%
  \BibitemOpen
  \bibfield  {author} {\bibinfo {author} {\bibfnamefont {M.-D.}\ \bibnamefont
  {Choi}}\ and\ \bibinfo {author} {\bibfnamefont {D.~W.}\ \bibnamefont
  {Kribs}},\ }\href@noop {} {\bibfield  {journal} {\bibinfo  {journal}
  {Physical review letters}\ }\textbf {\bibinfo {volume} {96}},\ \bibinfo
  {pages} {050501} (\bibinfo {year} {2006})}\BibitemShut {NoStop}%
\bibitem [{\citenamefont {Blume-Kohout}\ \emph {et~al.}(2010)\citenamefont
  {Blume-Kohout}, \citenamefont {Ng}, \citenamefont {Poulin},\ and\
  \citenamefont {Viola}}]{blume2010information}%
  \BibitemOpen
  \bibfield  {author} {\bibinfo {author} {\bibfnamefont {R.}~\bibnamefont
  {Blume-Kohout}}, \bibinfo {author} {\bibfnamefont {H.~K.}\ \bibnamefont
  {Ng}}, \bibinfo {author} {\bibfnamefont {D.}~\bibnamefont {Poulin}}, \ and\
  \bibinfo {author} {\bibfnamefont {L.}~\bibnamefont {Viola}},\ }\href@noop {}
  {\bibfield  {journal} {\bibinfo  {journal} {Physical Review A}\ }\textbf
  {\bibinfo {volume} {82}},\ \bibinfo {pages} {062306} (\bibinfo {year}
  {2010})}\BibitemShut {NoStop}%
\bibitem [{\citenamefont {B{\'e}ny}\ \emph {et~al.}(2007)\citenamefont
  {B{\'e}ny}, \citenamefont {Kempf},\ and\ \citenamefont
  {Kribs}}]{beny2007generalization}%
  \BibitemOpen
  \bibfield  {author} {\bibinfo {author} {\bibfnamefont {C.}~\bibnamefont
  {B{\'e}ny}}, \bibinfo {author} {\bibfnamefont {A.}~\bibnamefont {Kempf}}, \
  and\ \bibinfo {author} {\bibfnamefont {D.~W.}\ \bibnamefont {Kribs}},\
  }\href@noop {} {\bibfield  {journal} {\bibinfo  {journal} {Physical review
  letters}\ }\textbf {\bibinfo {volume} {98}},\ \bibinfo {pages} {100502}
  (\bibinfo {year} {2007})}\BibitemShut {NoStop}%
\bibitem [{\citenamefont {Kribs}\ and\ \citenamefont
  {Spekkens}(2006)}]{kribs2006quantum}%
  \BibitemOpen
  \bibfield  {author} {\bibinfo {author} {\bibfnamefont {D.~W.}\ \bibnamefont
  {Kribs}}\ and\ \bibinfo {author} {\bibfnamefont {R.~W.}\ \bibnamefont
  {Spekkens}},\ }\href@noop {} {\bibfield  {journal} {\bibinfo  {journal}
  {Physical Review A}\ }\textbf {\bibinfo {volume} {74}},\ \bibinfo {pages}
  {042329} (\bibinfo {year} {2006})}\BibitemShut {NoStop}%
\bibitem [{\citenamefont {Kribs}\ \emph
  {et~al.}(2005{\natexlab{a}})\citenamefont {Kribs}, \citenamefont {Laflamme},\
  and\ \citenamefont {Poulin}}]{kribs2005unified}%
  \BibitemOpen
  \bibfield  {author} {\bibinfo {author} {\bibfnamefont {D.}~\bibnamefont
  {Kribs}}, \bibinfo {author} {\bibfnamefont {R.}~\bibnamefont {Laflamme}}, \
  and\ \bibinfo {author} {\bibfnamefont {D.}~\bibnamefont {Poulin}},\
  }\href@noop {} {\bibfield  {journal} {\bibinfo  {journal} {Physical review
  letters}\ }\textbf {\bibinfo {volume} {94}},\ \bibinfo {pages} {180501}
  (\bibinfo {year} {2005}{\natexlab{a}})}\BibitemShut {NoStop}%
\bibitem [{\citenamefont {Knill}(2006)}]{knill2006protected}%
  \BibitemOpen
  \bibfield  {author} {\bibinfo {author} {\bibfnamefont {E.}~\bibnamefont
  {Knill}},\ }\href@noop {} {\bibfield  {journal} {\bibinfo  {journal}
  {Physical Review A}\ }\textbf {\bibinfo {volume} {74}},\ \bibinfo {pages}
  {042301} (\bibinfo {year} {2006})}\BibitemShut {NoStop}%
\bibitem [{\citenamefont {Wang}\ \emph {et~al.}(2013)\citenamefont {Wang},
  \citenamefont {Byrd},\ and\ \citenamefont {Jacobs}}]{wang2013numerical}%
  \BibitemOpen
  \bibfield  {author} {\bibinfo {author} {\bibfnamefont {X.}~\bibnamefont
  {Wang}}, \bibinfo {author} {\bibfnamefont {M.}~\bibnamefont {Byrd}}, \ and\
  \bibinfo {author} {\bibfnamefont {K.}~\bibnamefont {Jacobs}},\ }\href@noop {}
  {\bibfield  {journal} {\bibinfo  {journal} {Physical Review A}\ }\textbf
  {\bibinfo {volume} {87}},\ \bibinfo {pages} {012338} (\bibinfo {year}
  {2013})}\BibitemShut {NoStop}%
\bibitem [{\citenamefont {Shabani}\ and\ \citenamefont
  {Lidar}(2005)}]{shabani2005theory}%
  \BibitemOpen
  \bibfield  {author} {\bibinfo {author} {\bibfnamefont {A.}~\bibnamefont
  {Shabani}}\ and\ \bibinfo {author} {\bibfnamefont {D.~A.}\ \bibnamefont
  {Lidar}},\ }\href@noop {} {\bibfield  {journal} {\bibinfo  {journal}
  {Physical Review A}\ }\textbf {\bibinfo {volume} {72}},\ \bibinfo {pages}
  {042303} (\bibinfo {year} {2005})}\BibitemShut {NoStop}%
\bibitem [{\citenamefont {Wang}\ \emph {et~al.}(2016)\citenamefont {Wang},
  \citenamefont {Byrd},\ and\ \citenamefont {Jacobs}}]{wang2016minimal}%
  \BibitemOpen
  \bibfield  {author} {\bibinfo {author} {\bibfnamefont {X.}~\bibnamefont
  {Wang}}, \bibinfo {author} {\bibfnamefont {M.}~\bibnamefont {Byrd}}, \ and\
  \bibinfo {author} {\bibfnamefont {K.}~\bibnamefont {Jacobs}},\ }\href@noop {}
  {\bibfield  {journal} {\bibinfo  {journal} {Physical review letters}\
  }\textbf {\bibinfo {volume} {116}},\ \bibinfo {pages} {090404} (\bibinfo
  {year} {2016})}\BibitemShut {NoStop}%
\bibitem [{\citenamefont {Cirac}\ \emph {et~al.}(2017)\citenamefont {Cirac},
  \citenamefont {Perez-Garcia}, \citenamefont {Schuch},\ and\ \citenamefont
  {Verstraete}}]{cirac2017matrix}%
  \BibitemOpen
  \bibfield  {author} {\bibinfo {author} {\bibfnamefont {J.~I.}\ \bibnamefont
  {Cirac}}, \bibinfo {author} {\bibfnamefont {D.}~\bibnamefont {Perez-Garcia}},
  \bibinfo {author} {\bibfnamefont {N.}~\bibnamefont {Schuch}}, \ and\ \bibinfo
  {author} {\bibfnamefont {F.}~\bibnamefont {Verstraete}},\ }\href@noop {}
  {\bibfield  {journal} {\bibinfo  {journal} {Annals of Physics}\ }\textbf
  {\bibinfo {volume} {378}},\ \bibinfo {pages} {100} (\bibinfo {year}
  {2017})}\BibitemShut {NoStop}%
\bibitem [{\citenamefont {Cuevas}\ \emph {et~al.}(2017)\citenamefont {Cuevas},
  \citenamefont {Cirac}, \citenamefont {Schuch},\ and\ \citenamefont
  {Perez-Garcia}}]{cuevas2017irreducible}%
  \BibitemOpen
  \bibfield  {author} {\bibinfo {author} {\bibfnamefont {G.~D.~l.}\
  \bibnamefont {Cuevas}}, \bibinfo {author} {\bibfnamefont {J.~I.}\
  \bibnamefont {Cirac}}, \bibinfo {author} {\bibfnamefont {N.}~\bibnamefont
  {Schuch}}, \ and\ \bibinfo {author} {\bibfnamefont {D.}~\bibnamefont
  {Perez-Garcia}},\ }\href@noop {} {\bibfield  {journal} {\bibinfo  {journal}
  {arXiv preprint arXiv:1708.00029}\ } (\bibinfo {year} {2017})}\BibitemShut
  {NoStop}%
\bibitem [{\citenamefont {Kribs}\ \emph
  {et~al.}(2005{\natexlab{b}})\citenamefont {Kribs}, \citenamefont {Laflamme},
  \citenamefont {Poulin},\ and\ \citenamefont {Lesosky}}]{kribs2005operator}%
  \BibitemOpen
  \bibfield  {author} {\bibinfo {author} {\bibfnamefont {D.~W.}\ \bibnamefont
  {Kribs}}, \bibinfo {author} {\bibfnamefont {R.}~\bibnamefont {Laflamme}},
  \bibinfo {author} {\bibfnamefont {D.}~\bibnamefont {Poulin}}, \ and\ \bibinfo
  {author} {\bibfnamefont {M.}~\bibnamefont {Lesosky}},\ }\href@noop {}
  {\bibfield  {journal} {\bibinfo  {journal} {arXiv preprint quant-ph/0504189}\
  } (\bibinfo {year} {2005}{\natexlab{b}})}\BibitemShut {NoStop}%
\bibitem [{\citenamefont {Kuperberg}(2003)}]{kuperberg2003capacity}%
  \BibitemOpen
  \bibfield  {author} {\bibinfo {author} {\bibfnamefont {G.}~\bibnamefont
  {Kuperberg}},\ }\href@noop {} {\bibfield  {journal} {\bibinfo  {journal}
  {IEEE Transactions on Information Theory}\ }\textbf {\bibinfo {volume}
  {49}},\ \bibinfo {pages} {1465} (\bibinfo {year} {2003})}\BibitemShut
  {NoStop}%
\bibitem [{\citenamefont {Wolf}(2012)}]{wolf2012quantum}%
  \BibitemOpen
  \bibfield  {author} {\bibinfo {author} {\bibfnamefont {M.~M.}\ \bibnamefont
  {Wolf}},\ }\href@noop {} {\bibfield  {journal} {\bibinfo  {journal} {Lecture
  notes available at http://www-m5. ma. tum. de/foswiki/pub M}\ }\textbf
  {\bibinfo {volume} {5}} (\bibinfo {year} {2012})}\BibitemShut {NoStop}%
\bibitem [{\citenamefont {Blume-Kohout}\ \emph {et~al.}(2008)\citenamefont
  {Blume-Kohout}, \citenamefont {Ng}, \citenamefont {Poulin},\ and\
  \citenamefont {Viola}}]{blume2008characterizing}%
  \BibitemOpen
  \bibfield  {author} {\bibinfo {author} {\bibfnamefont {R.}~\bibnamefont
  {Blume-Kohout}}, \bibinfo {author} {\bibfnamefont {H.~K.}\ \bibnamefont
  {Ng}}, \bibinfo {author} {\bibfnamefont {D.}~\bibnamefont {Poulin}}, \ and\
  \bibinfo {author} {\bibfnamefont {L.}~\bibnamefont {Viola}},\ }\href@noop {}
  {\bibfield  {journal} {\bibinfo  {journal} {Physical review letters}\
  }\textbf {\bibinfo {volume} {100}},\ \bibinfo {pages} {030501} (\bibinfo
  {year} {2008})}\BibitemShut {NoStop}%
\bibitem [{\citenamefont {Guan}\ \emph {et~al.}(2016)\citenamefont {Guan},
  \citenamefont {Feng},\ and\ \citenamefont {Ying}}]{guan2016decomposition}%
  \BibitemOpen
  \bibfield  {author} {\bibinfo {author} {\bibfnamefont {J.}~\bibnamefont
  {Guan}}, \bibinfo {author} {\bibfnamefont {Y.}~\bibnamefont {Feng}}, \ and\
  \bibinfo {author} {\bibfnamefont {M.}~\bibnamefont {Ying}},\ }\href@noop {}
  {\bibfield  {journal} {\bibinfo  {journal} {arXiv preprint arXiv:1608.06024}\
  } (\bibinfo {year} {2016})}\BibitemShut {NoStop}%
\bibitem [{\citenamefont {Baumgartner}\ and\ \citenamefont
  {Narnhofer}(2012)}]{baumgartner2012structure}%
  \BibitemOpen
  \bibfield  {author} {\bibinfo {author} {\bibfnamefont {B.}~\bibnamefont
  {Baumgartner}}\ and\ \bibinfo {author} {\bibfnamefont {H.}~\bibnamefont
  {Narnhofer}},\ }\href@noop {} {\bibfield  {journal} {\bibinfo  {journal}
  {Reviews in Mathematical Physics}\ }\textbf {\bibinfo {volume} {24}},\
  \bibinfo {pages} {1250001} (\bibinfo {year} {2012})}\BibitemShut {NoStop}%
\bibitem [{\citenamefont {Ticozzi}\ and\ \citenamefont
  {Viola}(2008)}]{ticozzi2008quantum}%
  \BibitemOpen
  \bibfield  {author} {\bibinfo {author} {\bibfnamefont {F.}~\bibnamefont
  {Ticozzi}}\ and\ \bibinfo {author} {\bibfnamefont {L.}~\bibnamefont
  {Viola}},\ }\href@noop {} {\bibfield  {journal} {\bibinfo  {journal} {IEEE
  Transactions on Automatic Control}\ }\textbf {\bibinfo {volume} {53}},\
  \bibinfo {pages} {2048} (\bibinfo {year} {2008})}\BibitemShut {NoStop}%
\end{thebibliography}%

\end{document}